\documentclass[12pt]{article}
\usepackage[dvips]{graphicx}
\usepackage{latexsym}
\usepackage[abs]{overpic}
\usepackage{wrapfig}
\usepackage{amssymb}
\usepackage{amsmath}
\usepackage{color}
\usepackage{bm}

\input epsf
\newcommand{\R}{R}

\newcommand{\skyp}[1]{}

\def\Z {\bb{Z}}
\def\R {\bb{R}}

\def\rem#1{}
\def\C{{\mathbb{C}}} % \C=\mathbb{C}
\def\R{{\mathbb{R}}} % \R=\mathbb{R}
\def\Z{{\mathbb{Z}}} % \Z=\mathbb{Z}

\def\ep{{\epsilon}}

\def\bpsi{{\bar{\psi}}}
\def\bphi{{\bar{\phi}}}

\def\Tr{{\mathrm{Tr}}}

\def\rem#1{}

\newcommand\non{\nonumber \\}
\newcommand{\bel}{\begin{eqnarray}}
\newcommand{\ee}{\end{eqnarray}}

\makeatletter
\@addtoreset{equation}{section}
\makeatother
\begin{document}

\begin{titlepage}

\bigskip
\hfill\vbox{\baselineskip12pt
\hbox{}
\hbox{}
}
\bigskip\bigskip\bigskip\bigskip\bigskip\bigskip\bigskip\bigskip

\begin{center}
\Large{ \bf
Gauge/Bethe correspondence  on $S^1 \times \Sigma_h$ 

and 

index over moduli space
}
\end{center}
\bigskip
\bigskip

\bigskip
\bigskip
\centerline{ \large Satoshi Okuda$^{\clubsuit}$ and Yutaka Yoshida$^{\heartsuit}$}
\bigskip
\medskip
\centerline{$^{\clubsuit}$\it INFN, Sezione di Firenze and Dipartimento di Fisica e Astronomia}
%\centerline{\it  \ \ }
\centerline{ satoshi.okudaATfi.infn.it}
\bigskip
\bigskip
\centerline{$^{\heartsuit}$  \it School of Physics, Korea Institute for Advanced Study (KIAS)}
\centerline{\it  85 Hoegiro Dongdaemun-gu, Seoul, 130-722,  Korea }
\centerline{ yyyyosidaATgmail.com}
\bigskip
\bigskip

\begin{abstract}
We introduce two-types of topologically twisted Chern-Simons-matter theories on the direct product of circle and genus-$h$ 
Riemann surface ($S^1 \times \Sigma_h$). The partition functions of first model agrees with the partition functions of a generalizations of 
$G/G$ gauged WZW model. We also find that  correlation functions of Wilson loops in first type  Chern-Simons-matter theory
 coincide with correlation functions of $G$ elements in  the generalization of  $G/G$ gauged WZW  model.
The partition function of this model also has nice interpretations as norms of eigen states of Hamiltonian in the quantum integrable model (q-boson hopping model) 
and also as a geometric index over a particular moduli space.
In the second-type    Chern-Simons-matter theory, the partition function  is related to 
integration over moduli space of  Hitchin equation on Riemann surface.
\end{abstract}
\end{titlepage}

\newpage
\baselineskip=18pt

\tableofcontents
\section{Introduction}
$G$  Wesss-Zumino-Witten (WZW) model is a non-linear sigma model whose  target space is given by a Lie group $G$. 
In two dimensions, $G$ WZW model  has been  studied extensively in the context of  conformal field theory. %It found that many interesting structures. 
$G/G$ gauged WZW   model we concern  is obtained by gauging  Lie group $G$ in $G$ WZW model. 
  $G/G$ gauged WZW model also possesses many interesting structures. For examples:
\\
\\
1. The partition  function $\mathcal{Z}_{\text{GWZW}}(\Sigma_h)$ of $G/G$ gauged WZW model  on genus $h$ Riemann surface $\Sigma_h$      
is expressed by modular $S$-matrices \cite{Blau:1993tv, Gerasimov:1993ws} as
\bel
 \mathcal{Z}_{\text{GWZW}}(\Sigma_h)=\sum_{\mu} \frac{1}{S^{2h-2}_{0 \mu} }
\label{CSpar}
\ee
\\
2. $G/G$ gauged WZW model is a two dimensional topological quantum field theory \cite{Witten:1993xi}.
\\
\\
3. $G/G$ gauged WZW model possesses a hidden qunatum integrable structure \cite{Korff:2010, Okuda:2012nx}
\\
\\
4. The partition function  $\mathcal{Z}_{\text{GWZW}}(\Sigma_h)$ is positive  integer which is given by  an index over the moduli space of
$G$ flat connections on Riemann surface $\Sigma_h$ \cite{Witten:1991we}.
\\
\\
5.   $\mathcal{Z}_{\text{GWZW}}(\Sigma_h)$ 
is identical to the partition function $\mathcal{Z}_{\text{CS}}(S^1 \times \Sigma_h)$ of $G$ pure Chern-Simons theory on $S^1 \times \Sigma_h$:
\bel
 \mathcal{Z}_{\text{GWZW}}(\Sigma_h)=\mathcal{Z}_{\text{CS}}(S^1 \times \Sigma_h).
\ee
In our previous work \cite{Okuda:2013fea}, we constructed a one parameter generalization of $G/G$ gauged WZW model  ($G/G$ gauged WZW-matter model) and 
evaluated its partition function and correlation functions. 
 We showed that $G/G$ gauged WZW-matter model possesses following similar properties to $G/G$ gauged WZW model:
\\
\\
1'. The partition  function $\mathcal{Z}_{\text{GWZWM}}(\Sigma_h)$ of $G/G$ gauged WZW model  on genus $h$ Riemann surface $\Sigma_h$      
is expressed by one-parameter dedormation of   modular $S$-matrices as
\bel
 \mathcal{Z}_{\text{GWZWM}}(\Sigma_h)=\sum_{\mu} \frac{1}{S^{2h-2}_{0 \mu} (t) }.
\ee
\\
2'. $G/G$ gauged WZW-matter   model is a two dimensional topological quantum field theory.
\\
\\
3'. $G/G$ WZW-matter model model possesses a hidden quantum integrable structure. 
\\
\\
From an analogy of gauged WZW model,  we  also suggested following conjectures:
\\
\\
4'. The partition function  $\mathcal{Z}_{\text{GWZWM}}(\Sigma_h)$ is related to a partition function of topological twisted Chern-Simon-matter theory on $S^1 \times \Sigma_h$
constructed in \cite{Ohta:2012ev}.
\\
\\
5'. The partition function of $G/G$ WZW-matter  model  is related to a geometric index over moduli space \cite{Teleman:2003}.
\\
\\
One purpose of this paper is to reveal the relation between $G/G$ gauged WZW-matter model on $\Sigma_h$ 
and twisted Chern-Simons-matter theory on $S^1 \times \Sigma_h $. 
The second motivation of this paper is to relate the twisted Chern-Simons-matter theories to integration over  moduli spaces defined particular 
differential equations. 
In the case of pure Chern-Simons theory, it is not difficult to see the path integral localize to the moduli space of flat connections. 
But, in the previous work,  it was not clear  how  the path integral of $G/G$ gauged WZW-matter model  localized to moduli spaces.    
The third motivation is Gauge/Bethe correspondence discovered in \cite{Moore:1997dj, Nekrasov:2009uh}. So far, it has not been  well studied that what types of objects in the qunatum integrable model correspond to the partition functions or correlation functions in the gauge theory side.
  In \cite{Okuda:2013fea}, we revealed that
the  partition function (correlation functions) of $G/G$ gauged WZW-matter is related to the norms of eigen states of Hamiltonian ( the correlation functions of conserved charges ) 
in the quantum integrable model, respectively. In this case, the corresponding quantum integrable model is  $q$-boson hopping model  \cite{Kulish:1991bk}. 
From the above conjectures, we expect  that the partition function and correlation functions in the twisted Chern-Simons-matter theory are also 
related to norms of eigen states and correlation functions in the  quantum integrable model.  

This paper is organized as follows. In section \ref{section2}, we introduce two types of twisted Chern-Simon-matter theories. 
Both of two models can be embedded into $S^1$-uplift of A-twisted Chern-Simons-matter theories. The difference between
 two models is R-charge assignment before twisting.
In subsection \ref{model1},  we introduce a twisted Chern-Simons-matter theory on $S^1 \times \Sigma_h$ which is related to
$G/G$ gauged WZW-matter model on $\Sigma_h$. 
We evaluate the partition function and show the equivalence between the twisted Chern-Simons-matter theory and $G/G$ gauged WZW-matter model.
This means that Chern-Simons-matter theory also possesses quantum integrable structure and TQFT structure. 
In subsection \ref{model2}, we introduce another  twisted Chern-Simons-matter theory.
In subsection \ref{correlation}, we generalize the correspondence between the partition functions   to  correlation functions. 
Wilson loop  correlation   functions in the first type twisted Chern-Simons-matter theory correspond to correlation functions of $G$-elements in the  $G/G$ gauged
WZW-matter models. 
In section \ref{secindex}, we study the interpretation of the partition functions of  two Chern-Simons-matter theories in terms of moduli space.  
The last section is devoted to summary.
 
%%%%%%%%%%%%%%%%%%%%%%%%%%%%%%%%%%%%%%%%%%%%%%%%%%%%%%%%%%%%%%%%%%
\section{Twisted Chern-Simons matter theory on $S^1 \times \Sigma_h$}
\label{section2}

In this section we evaluate twisted Chern-Simons-Matter theories  on the direct product of circle and genus-$h$ Riemann surface $S^1 \times \Sigma_h$.
The model  consists of Chern-Simons term and the topologically twisted matter sectors.  The  twist we introduce  is $S^1$-uplift  of A-twist in two dimensions \cite{Ohta:2012ev}.    
We take local coordinate of $S^1 \times \Sigma_h$ as $(x^0, x^1, x^2)=(t, z, \bar{z})$ and take circumference of $S^1$ as $2\pi$.

%%%%%%%%%%%%%%%%%%%%%%%%%%%%%%%
First we consider  Chern-Simons term.
In order to introduce matter multiplets in the BRST exact manner. We introduce  adjoint 1-form valued fermion $\lambda_i,  (i=1,2)$ and scalar Grassmann odd  symmetry $Q$ (BRST symmetry). When we do not couple Chern-Simons theory with
twisted matters, $\lambda_i$ are auxiliary fields and do not affect theory explicitly. 
   
The  $Q$-transformation is defined by
\bel
&&Q A_{j} = \lambda_{j}, \quad  Q \lambda_j = F_{t j }, \quad  (j=1,2) ,\non
&&Q A_{t} = 0.
\ee
 
The Chern-Simons action is defined by
\bel
S_\text{CS}[A] &&= \frac{i k}{4\pi} \int_{S^1 \times \Sigma_h} 
\Tr\left[
A \wedge dA +\frac{2i}{3}A \wedge A \wedge A
\right] \non
&&=\frac{i k}{4\pi} \int_{S^1 \times \Sigma_h} d^3 x
\ep^{\mu \nu \rho} \Tr\left[
A_{\mu} \partial_{\nu} A_{\rho} +\frac{2i}{3} A_{\mu}  A_{\nu}  A_{\rho}
\right].
\label{bosonic CS}
\ee
The $Q$-closed completion of Chern-Simons action is given by
\bel
S_\text{coh}[A,\lambda] = S_\text{CS}[A] 
+ \frac{i k}{4\pi} \int_{S^1 \times \Sigma_h} 
d^3 x \Tr\left[
 \ep^{ij} \lambda_i \lambda_j 
\right].
\ee

The bosonic part of $S_\text{coh}$ is written as
\bel
S_\text{CS}[A] =
\frac{i k}{4\pi} \int_{S^1 \times \Sigma_h}
\ep^{ij}  \mathrm{Tr} \left[   A_t F_{ij}  - A_{i} \partial_t A_{j}
\right],
\label{CohCSterm}
\ee
Integrating out a gauge field $A_t$ in the pure Chern-Simons action  imposes a  delta functional constraint $\delta (F_{12}) $ and the partition function is interpreted as partition function of quantum mechanical system whose target space is the moduli space of  the  flat connection condition $F_{1 2}=0$ on the Riemann surface. Then the  quantization of Chern-Simons theory on $S^1 \times \Sigma_h$ is the  quantization problem of moduli space of flat connection on Riemann surface $\Sigma_h$.
We will  see the similar interpretation is possible for twisted Chern-Simons-Matter theories in section \ref{secindex}.

Next, we  couple Chern-Simons theory to matters fields. We are interested in  one-parameter deformation of Chern-Simons theory which are equivalent to one-parameter deformation of  $G/G$ gauged WZW model constructed in \cite{Okuda:2013fea} and \cite{Gerasimov:2006zt}. 

%%%%%%%%%%%%%%%%%%%%%%%%%%%%%%%%%%%%%%%%%
%%%%%%%%%%%%%%%%%%%%%%%%%%%%%%%%%%
%%%%%%%%%%%%%%%%%%%%%%%%%%%%%%%%%%%%
\subsection{The model 1}
\label{model1}

First we consider a deformation of  Chern-Simons theory which is equivalent to $G/G$ gauged WZW-matter model introduced  in \cite{Okuda:2013fea}.
The matter fields we introduce are following:
\begin{itemize}
\item
The Grassmann even matter: $(\phi, \overline{\phi}, Y_{\bar{z}}, Y_{{z}})$
\item
The Grassmann odd matter $(\psi, \overline{\psi}, \chi_{\bar{z}}, \chi_{{z}})$
\end{itemize}
$\phi$ and $\overline{\phi}$ are complex scalar,
 $Y_{{z}}, \chi_{{z}}$  are $(1,0) $-form   and $Y_{\bar{z}},  \chi_{\bar{z}}$ are $(0,1) $-form along Riemann surface, respectively. 
All the matter fields belong to complexified adjoint representation of the Lie algebra $\mathfrak{g}$.
The BRST transformation is defined by 
\bel
&&Q \phi = \psi, \quad  Q \psi =  D_t  \phi + m \phi , \non
&&Q Y_{\bar{z}} =D_t \chi_{\bar{ z}} - m \chi_{\bar{z}}, \quad Q \chi_{\bar{z}} =Y_{\bar{z}}, \non
&&Q \bphi = \bpsi, \quad  Q \bpsi = D_t \bphi - m \bphi , \non
&&Q Y_{{z}} =D_t \chi_{{ z}} + m  \chi_{z}, \quad Q \chi_{{z}} =Y_{{z}}.
\label{Atwist}
\ee
$D_{t} \phi := \partial_t \phi +i [A_t, \phi]$ and so on.
Here $m$ is a equivariant mass parameter (real mass)  which regularizes the flat direction. When we perform the dimensional reduction along 
 the $S^1$-direction. Then (\ref{Atwist}) reduces to the BRST transformation of  A-twisted chiral multiplet on the Riemann surface.   
We introduce the action for matter fields as  $Q$-exact matter: 
\bel
S_{\text{mat}1} && = \int_{S^1 \times \Sigma_h} Q \mathrm{Tr} \Bigl[   \overline{\phi}  \psi 
 + g^{\bar{z} z } ( Y_{\bar{z}}  \chi_{{z}} %+\chi_{\bar{z}} Y_{{z}} 
 ) \Bigr] \non
&&=\int_{S^1 \times \Sigma_h} \mathrm{Tr} \Bigl[   \overline{\phi}  (D_t \phi + m \phi ) + \overline{\psi} \psi 
+ g^{z \bar{z}} Y_{{z}}  Y_{\bar{z}} 
   - g^{z \bar{z}}  \chi_{\bar{z}} (D_t \chi_{{z}}  + m {\chi}_z )      \Bigr].
\label{matteraction1}   
\ee
The action is again invariant under the $Q$-transformation.
The partition function of twisted  Chern-Simons-Matter theory is defined by
\bel
\mathcal{Z}_{\text{CSM1}} (S^1 \times \Sigma_h) 
=\frac{1}{\text{Vol}(\mathcal{G})}\int \mathcal{D}^3 A_{\mu} \mathcal{D}^2 \lambda_{z} \mathcal{D}^2 \phi   \mathcal{D}^2 \psi \mathcal{D}^2 Y_{i} \mathcal{D}^2 \chi_{i}
  e^{S_{\text{coh}}- S_{\text{mat}}}. 
  \label{eff1}
\ee 
Here $\mathcal{G}$ is the gauge transformation group of $G$ and $\text{Vol} (\mathcal{G})$ is it volume. 
We evaluate the partition function. 
We take the Cartan-Weyl basis $(H_a, E_{\alpha})$ of the Lie algebra $\mathfrak{g}$ of gauge group $G$. 
The commutation relations in the Cartan-Weyl basis are
\begin{eqnarray}
[H^a,H^b] =0,\quad
[H^a,E^{\alpha}] = {\alpha}^a E^{\alpha}
\end{eqnarray}
%and
%\begin{eqnarray}
%[E^{\alpha},E^{\beta}] 
% &=& N_{\alpha,\beta}E^{\alpha+ \beta}, \quad \mathrm{if}~\alpha+\beta \in \Delta\nonumber\\
% &=& \frac{2}{|\alpha|^2}\alpha\cdot H,\quad \mathrm{if}~\alpha = -\beta\nonumber\\
%&=& 0\quad \mathrm{otherwise}
%\end{eqnarray}
%where $\alpha=(\alpha^1,\cdots,\alpha^N)$ is root vector and $N$ is rank of Lie algebra.
Trace is defined as
\begin{eqnarray}
\Tr(H^a H^b) = \delta_{ab}, 
\quad \Tr(E^{\alpha}E^{\beta}) = \frac{2}{|\alpha|^2} \delta^{\alpha+\beta,0}.
\end{eqnarray}
Let $\Psi$ is a generic Lie algebra valued field. Then,
$\Psi$ is expanded by the Cartan-Weyl basis as follows.
\bel
\Psi=\sum_{a} \Psi^a H^a + \sum_{\alpha \neq 0 } \Psi^{\alpha} E^{\alpha}. 
\ee
From now on, we assume gauge group is $U(N)$. 
We take following gauge \cite{Blau:2006gh}
\bel
\partial_t A^{a}_t =0 \quad  \text{and} \quad A^{\alpha}_t=0. 
\label{gaugefix}
\ee
The first condition in (\ref{gaugefix}) requires the gauge field along time direction depends only on the coordinates on Riemann surface.
Then ghost action associated to the above gauge fixing condition is given by
\bel
S_{\text{gh}}=\int_{S^1 \times \Sigma_h} \mathrm{Tr} (\bar{c} \partial_t c +i \bar{c} [A^a_t H^a, c]).
\ee
Next we expand the three dimensional fields by Kaluza-Klein (KK) modes along the $S^1$-direction as
\bel
\Psi(t, z,\bar{z})=\sum_{n \in \Z} \Psi_n (z,\bar{z}) e^{i n t}. 
\ee
By integrating out $(c, \bar{c})$, the one-loop determinant of ghost action is formally written as
\bel
Z^{1\text{-loop}}_{\text{ghost}}&&=\prod_{n \in \Z} \prod_{\alpha \neq 0} \mathrm{Det}_{\Omega^{(0,0)} }  \left(i n  +i  \alpha (A_t) \right) \non
&&=\prod_{\alpha \neq 0} \mathrm{Det}_{\Omega^{(0,0)} }   \left( 1-e^{2 \pi i \alpha (A_t)} \right) 
\label{ghostdet}
\ee
Here $n \in \Z$ expresses the KK modes along the $S^1$-direction and $\mathrm{Det}_{\Omega^{(n,m)}}$ is functional determinant
associated to the section of $(n,m)$-form on the Riemann surface.  
We can also integrate out  the gauge field along Riemann surface $A^{\alpha}_{i}$, then we  obtain the functional determinant:  
\bel
Z^{1\text{-loop}}_{A_{i}}&&=\prod_{n \in \Z} \prod_{\alpha \neq 0} \mathrm{Det}_{\Omega^{(1,0)} }  \left(n i  + i \alpha(A_t) \right)^{-1} 
 \non
&&=\prod_{\alpha > 0} \mathrm{Det}_{\Omega^{(1,0)} }   \left(1- e^{2 \pi i \alpha (A_t) }\right)^{-1} \mathrm{Det}_{\Omega^{(0,1)} }   \left(1- e^{-2 \pi i \alpha (A_t) }\right)^{-1} 
\label{gaugedet}
\ee    
Each functional determinant  (\ref{gaugedet}) and (\ref{ghostdet}) is infinite dimensional and not well-defined, but 
the ratio between these determinants are evaluated by using heat kernel regularization as
\bel
&&\prod_{\alpha > 0}
 \frac{\mathrm{Det}_{\Omega^{(0,0)} }   \left( 1-e^{2 \pi i \alpha (A_t)} \right) }
{\mathrm{Det}_{\Omega^{(1,0)} } \left( 1-e^{2 \pi i \alpha (A_t)} \right) }  
 \frac{\mathrm{Det}_{\Omega^{(0,0)} }   \left( 1-e^{-2 \pi i \alpha (A_t)} \right) }
{\mathrm{Det}_{\Omega^{(0,1)} } \left( 1-e^{-2 \pi i \alpha (A_t)} \right) } 
 \non
&&~~=\prod_{\alpha \neq 0} \exp \Bigl\{  \frac{1}{8\pi} \int_{\Sigma_h} R \log \left( 1-e^{2 \pi i \alpha (A_t)} \right)   
+\frac{1}{2\pi} \int_{\Sigma_h}  \alpha_a F^a \log \left( 1-e^{2 \pi i \alpha (A_t)} \right)  \Bigr\}. 
\label{detvec}
\ee
Here $R$ is the scalar curvature associated to the metric on Riemann surface and $F_a, (a=1, \cdots, \text{rank}(\mathfrak{g}) ) $ are 
field strengths of the Cartan part of gauge field. 

Next we evaluate the functional determinant from the matter fields.
By integrating out $(\overline{\phi}, \phi)$ and $(\chi_{{z}}, \chi_{\bar{z}})$, the ratio of functional determinant are again evaluated by 
heat kernel regularization as 
\bel
&&\frac{\mathrm{Det}_{\chi} (\partial_t +i \alpha(A_t) - m)}{\mathrm{Det}_{\phi} (\partial_t +i \alpha(A_t) - m)}
= (1-e^{-2\pi m})^{N(h-1)} \non 
&& ~~~~~\times \prod_{\alpha \neq 0} \exp \Bigl\{ - \frac{1}{8\pi} \int_{\Sigma_h} R \log   (1- e^{ 2 \pi (i \alpha (A_t) - m)}  )  
-\frac{1}{2\pi} \int_{\Sigma_h}  \alpha_a F^a \log { (1- e^{ 2 \pi (i \alpha (A_t) - m)}  ) }  \Bigr\}. \non
\label{detmatter1}
\ee
From (\ref{detvec}) and(\ref{detmatter1})
The partition function (\ref{eff1}) can be written as 
\bel
\mathcal{Z}_{\text{CSM1}}(S^1 \times \Sigma_h) &&=(1-e^{-2\pi m})^{N(h-1)}
 \int   \mathcal{D} A^{a 3} \mathcal{D} \lambda^{a 2}  
\prod_{a \neq b } \exp \Bigl\{  \frac{1}{8\pi} \int_{\Sigma_h} R \log \left( \frac{1-e^{(2 \pi i (A^a_t -A^b_t )}} {1 -e^{  2 \pi i (A^a_t -A^b_t +im)}  } \right) \Bigr\} 
\non &&~~~~~~\times
  \exp \Bigl[ \int_{\Sigma_h}  
  \sum_{a=1}^N i  \beta (A^a_t) {F^a}  - \int_{S^1 \times \Sigma_h }\frac{ik}{4 \pi } \ep^{ij} \lambda^a_i  \lambda^a_j \Bigr] 
\label{effaction1}
\ee
with 
\bel
\beta^a (x):=k x^a -\sum_{b=1 \atop b \neq a} \frac{i}{2\pi} \log \frac{e^{2\pi i x^a} -e^{-2 \pi m} e^{2\pi i x^b}}{e^{-2\pi m} e^{2\pi i x^a} - e^{2\pi i x^b}}
\ee
In  \cite{Okuda:2013fea}, we conjectured  the existence of twisted Chern-Simons-Matter theory whose partition function 
 is identical to $G/G$ gauged WZW-matter model.  Especially, we expected that the 
equivariant coupling constant which regularize th flat direction  in the gauged   WZW-matter model
$t$ (not time coordinate)  is    related to mass parameter in the corresponding Chern-Simons-Matter theory.
In fact (\ref{effaction1}) is precisely same as the Abelianized effective action of $G/G$ gauged WZW-matter model in \cite{Okuda:2013fea} with the identification $e^{-2 \pi m}=t$.
Therefore we find that partition function of the Chern-Simons-matter  (\ref{partition1}) agrees with the the partition function of $U(N)/U(N)$ gauged WZW-matter model $\mathcal{Z}_{\text{GWZWN}}$ on genus $h$ Riemann surface:
\bel
 \mathcal{Z}_{\text{CSM}1}(S^1 \times \Sigma_h) = \mathcal{Z}_{\text{GWZWM}} (\Sigma_h).
\label{corre1}
\ee 
The evaluation of the path integral of $\mathcal{D} A^3_a \mathcal{D} \lambda^2_a$  is same as that of 
partition function of $G/G$ gauged WZW-matter model. So we only briefly mention on the derivation.  
We decompose the Abelianized field strength to harmonic part $F^{a}_{(0)}$ and a exterior derivative of some one-form part $\tilde{A}^a$ as
 $F^a =F^{a}_{(0)}+d \tilde{A}^a$ 
Here  $F^a =F^{a}_{(0)}$ satisfies Dirac quantization condition
\bel
k^a=\frac{1}{2\pi} \int_{\Sigma_h} F^{a}_{(0)},  \quad k^a \in \Z.
\ee
The path integral of $\tilde{A}^a$ with the partial integration $d \tilde{A^a} (\cdots)= \tilde{A^a} d (\cdots) $ imposes $A^a_t$ to constant configuration $\varphi^a$.
By using Poisson resummation formula, the summation over magnetic charges $k_a$,  
imposes the field configuration $\varphi^a$ to satisfy the solutions 
\bel
\beta^a_1 (\varphi)=
  n^a, \quad n^a \in \Z.
\label{bethe}
\ee
This equation (\ref{bethe}) is precisely same as the logarithmic form of Bethe Ansatz equation of $q$-boson hopping model.
Taking account into BRST completion of gaugino harmonic modes, the partition function is evaluated as
\bel
\mathcal{Z}_{\text{CSM1}}(S^1 \times \Sigma_h) 
&&=
(1-e^{-2\pi m})^{N(h-1)} \non
&& \times \sum_{ \{ \varphi^a \}^{N}_{a=1} \in \mathcal{B}_1}  
  \left( \Bigl| \text{det}_{c,d} \left[  \frac{\partial \beta^c_1 (\varphi)}{ \partial \varphi^d } \right] \Bigr| \prod_{a \neq b }
 \frac{  e^{2\pi i \varphi_a} -e^{-2\pi m} e^{2\pi i \varphi_b} }
{e^{2\pi i \varphi_a} - e^{2\pi i \varphi_b}} \right)^{h-1} 
\label{partition1}
\ee
Here $\mathcal{B}_1$ is the set of  the solutions of the Bethe Ansatz equations (\ref{bethe}). A overall $\varphi^a$ independent ambiguity exists, but  
such a ambiguity   does not affect the correlation functions normalized by the partition function.  
In the the $G/G$ gauged WZW-matter model side, $\varphi^a$ comes from the constant configuration of diagonalized $G$-element 
 $g=e^{2\pi i \sum_a \varphi^a H^a}: \Sigma_h \to U (1)^N \subset U(N)$.

In the dictionary of Gauge/Bethe correspondence \cite{Nekrasov:2009uh}, 
the saddle point equation of effective twisted super potential of three dimensional $\mathcal{N}=2$ supersymmetric theory on $S^1 \times \R^2$ gives 
Bethe Ansatz equation in the corresponding quantum integrable system. As pointed out in \cite{Yoshida:2014ssa}, The saddle point equation of
 the twisted superpotential in $\mathcal{N}=2$ 
supersymmetric Chern-Simons-matter theory with an adjoint chiral multiplet give the Bethe Ansatz equation   of $q$-boson model. 
Actually, before the topologically twisting, the matter content  is same.

Next we mention about  the properties of twisted Chern-Simons-matter theory.
\begin{itemize}
\item
Gauge/Bethe correspondence

In the previous paper \cite{Okuda:2013fea}, we  showed that the partition function of $U(N)/U(N)$ gauged WZW-matter model $\mathcal{Z}_{\text{GWZWM}}$ is expressed as
norms of eigen states of Hamiltonian in the $q$-boson model:
\bel
\mathcal{Z}_{\text{GWZWM}}(\Sigma_h) &&=\sum_{\{ \varphi^a\}^N_{a=1} \in \mathcal{B}_1}
 \langle \psi_N (\{ e^{2\pi i \varphi^a} \}, t) | \psi_N (\{ e^{2 \pi i \varphi^a} \}, t)  \rangle^{h-1} \non
&&= \sum_{\{ \varphi^a\}^N_{a=1} \in \mathcal{B}_1} || \psi_N (\{  e^{2\pi i \varphi^a} \}, t)  ||^{2h-2}. 
\ee
Here $ | \psi_N ( \{ e^{2\pi i \varphi^a} \} , t)  \rangle$ are the eigen vectors of Hamiltonian of $q$-boson model in the $N$-particle sector and $\langle \psi_N ( \{ e^{2\pi i \varphi^a} \}, t) |$ is the dual vector of $| \psi_N ( \{ e^{2\pi i \varphi^a} \} , t)  \rangle$. In the $q$-boson model side, $t$ corresponds to the $q$-deformation parameter. 
The limit $t \to 0$ corresponds to the strong coupling limit. In this limit $q$-boson model reduces to phase model. 
In the gauged WZW-matter (Chern-Simons-matter model ) side, matter decouples from the WZW model (pure Chern-Simons theory) in the limit $t\to 0, (m \to \infty)$.
The gauge/Bethe correspondence between $G/G$ gauged WZW model (pure Chern-Simons theory) 
and  phase was studied in \cite{Okuda:2012nx}. Again the partition function of $G/G$ gauged WZW model on
 $ \Sigma_h$ is expressed by the norms of eigen states in the phase model.
\end{itemize}

\begin{itemize}
\item
Deformed Verlinde formula and TQFT structure

In \cite{Okuda:2013fea} we also showed that the partition function of gauged WZW-matter model  is expressed by one-parameter deformation of  modular $S$-matrices $S_{\mu \nu}(t), (\mu, \nu \in \mathcal{A}^{+}_{N,k})$ introduced by Korff \cite{Korff:2013rsa} and that the partition function $\mathcal{Z}_{\text{GWZWM}}$ or equivalently $\mathcal{Z}_{\text{CSM}1}$  can be constructed from axiom of two dimensional topological quantum field theory   (TQFT) \cite{Atiyah:1989vu, Segal:2002ei}. 
Here $ \mathcal{A}^{+}_{N,k}$ is a subset of the dominant integrable positive weights of $\mathfrak{gl}(N)$ and is identified with the space of the solution of 
$q$-boson Bethe Ansatz equation (\ref{bethe}). Especially $S_{0 \nu}(t)=|| \psi_N (\{ e^{2\pi i \varphi^a} \}, t)  ||^{-1}$. 
The $S_{\mu \nu}(t)$ can be regarded as one-parameter deformation of  modular $S$-matrices, Because $S_{\mu \nu}(t)$ 
reduces to modular $S$-matrices in $t \to 0, (m \to \infty)$.
From the equivalence (\ref{corre1}), the partition function of Chern-Simons-Matter theory is also expressed by deformed modular matrices as 
\bel
\mathcal{Z}_{\text{CSM}1} (S^1 \times \Sigma_h)&&=\sum_{\{ \varphi^a\}^N_{a=1} \in \mathcal{B}_1}
 \langle \psi_N (\{ e^{2\pi i \varphi^a} \}, t) | \psi_N (\{ e^{2 \pi i \varphi^a} \}, t)  \rangle^{h-1} \non
&& = \sum_{\mu \in \mathcal{A}^{+}_{N,k}} \frac{1}{S^{2h-2}_{0 \mu} (t)}
\label{CSmodular}
\ee 
In the limit $t \to 0$, (\ref{CSmodular}) reduces to (\ref{CSpar}).

We comment on the properties of partition function (\ref{CSmodular}).
Since $S_{\mu \nu}(t)$ is function of the roots of Bethe Ansatz equation,  
it is difficult to compute the each deformed modular matrix $S_{\mu \nu}(t)$ itself. But as we mention, (\ref{CSmodular}) is explicitly calculable 
by using axiom of 2d TQFT. The 2d TQFT structure we concern is identical to 
a finite dimensional commutative  Frobenius algebra constructed by Korff. \footnote{There is one to one correspondence between two dimensional topological quantum field theory and finite dimensional commutative  Frobenius algebra} Remarkably we found that the partition function of $U(N)/U(N)$ gauged WZW-matter model becomes a polynomial of $t$ for $h \ge 1$ with the integer coefficients and  series with the integer coefficients for $h=0$.   
The origin of integer valueness comes from properties of  deformed fusion coefficient introduced in \cite{Korff:2013rsa}:  
\bel
N^{\lambda}_{\mu \nu} (t):=\sum_{\sigma \in \mathcal{A}^+_{N,k}} \frac{S_{\mu \sigma}(t) S_{\nu \sigma}(t) S^{-1}_{\sigma \lambda}(t) }{S_{0 \sigma}(t)}.
\label{Verline}
\ee
$N^{\lambda}_{\mu \nu} (t)$ can be regarded as one-parameter deformation of Verline formula.
The important point is that Korff derived the explicit algorithm to compute $N^{\lambda}_{\mu \nu} (t)$ in terms of Bethe ansatz of $q$-boson and constructed 
the Frobenius algebra whose structure constant is given by $N^{\lambda}_{\mu \nu} (t)$. 
$N^{\lambda}_{\mu \nu} (t)$ become polynomials with integer coefficients. In the TQFT side, $N^{\lambda}_{\mu \nu} (t)$ corresponds to the data assigned to the pant 
(three punctured sphere). From the integer valueness of deformed fusion coefficients, we can show the integer valueness of  the partition function of 
$G/G$ gauged WZW-matter model. 
 
We suggest a brief explanation of the integer valuedness of partition function in terms of Hilbert space interpretation of Chern-Simons-Matter theory.
Recall that the partition function of  pure Chern-Simons theory on $S^1 \times \Sigma_h$ is 
expressed by trace over the Hilbert space on Riemann surface $\mathcal{H}(\Sigma_h)$.
\bel
\mathcal{Z}_{\text{CS}} (S^1 \times \Sigma_h)=\mathrm{Tr}_{\mathcal{H}_{\text{CS}}(\Sigma_h)} 1 .
\ee
It is known that  the Hilbert space $\mathcal{H}(\Sigma_h)$ is finite dimensional and   
same as the space of conformal block of WZW model on Riemann surface. Thus the partition agrees with the number of conformal blocks and integer.
On the other hand, the partition function of twisted Chern-Simons-Matter theory is interpreted as  the following index.   
\bel
\mathcal{Z}_{\text{CSM1}} (S^1 \times \Sigma_h)=\mathrm{Tr}_{\mathcal{H}_{\text{CSM}}(\Sigma_h)}  e^{-2\pi Q_{U(1)}m } .
\ee
Here $Q_{U(1)}$ is the charge associated to $U(1)$-rotation like this $(\overline{\phi}, \phi) \to (e^{-i \alpha } \overline{\phi}, e^{i \alpha } \phi)$.
So each  coefficient of the partition function should be an integer. But still integer valueness of partition function is rather  mysterious in 
the path integral formalism. We will explain its geometric origin in terms of index over moduli space in the section \ref{secindex}.
\end{itemize}

%%%%%%%%%%%%%%%%%%%%%%%%%%%%%%%%%%%%%%%%
%%%%%%%%%%%%%%%%%%%%%%%%%%%%%%%%%%%%%%%%
%%%%%%%%%%%%%%%%%%%%%%%%%%%%%%%%%%%%%%%%
\subsection{The  model 2 }
\label{model2}
Next we consider a deformation of  Chern-Simons theory which is related to a generalization of $G/G$ gauged WZW model introduced  in \cite{Gerasimov:2006zt}.
The matter fields  are 
\begin{itemize}
\item
The Grassmann even matter: $(\phi_{\bar{z}}, \phi_{z}, \overline{Y}, Y)$
\item
The Grassmann odd matter $(\psi_{\overline{z}}, \psi_{z}, \overline{\chi}, \chi)$
\end{itemize}
All the matter fields again belong to adjoint representation of the Lie algebra.
The BRST transformation is defined by 
\bel
&&Q \phi_{\bar{z}} = \psi_{\bar{z}}, \quad  Q \psi_{\bar{z}} = D_t  \phi_{\bar{z}}+ m \phi_{\bar{z}} , \non
&&Q Y =D_t \chi - m \chi, \quad Q \chi =Y, \non
&&Q \phi_{z} = \psi_{z}, \quad  Q \psi_{z} =  D_t \phi_{z} - m \phi_z, \non
&&Q \overline{Y} =D_t \overline{\chi} + m  \overline{\chi}, \quad Q \overline{\chi} =\overline{Y}.
\label{Atwist2}
\ee

Again we introduce the $Q$-exact matter action.
\bel
S_{\text{mat2}} && = 
\int_{S^1 \times \Sigma_h} Q \mathrm{Tr} \Bigl[  g^{z \bar{z}} {\phi}_{z}  \psi_{\bar{z}}  + \overline{Y}  \chi \Bigr] \non
&&=\int_{S^1 \times \Sigma_h} \mathrm{Tr} \Bigl[   g^{z \bar{z}} \phi_z  (D_t \phi_{\bar{z}}  + m \phi_{\bar{z}} ) + g^{z \bar{z}} \psi_z \psi_{\bar{z}}
+   \bar{Y} Y    +\overline{\chi}  (D_t \chi + m \chi)  \Bigr].
\ee 
If we replace $\phi_{\bar{z}}$ by $\phi$, $\psi_{\bar{z}}$ by $\psi$ and so on, 
the action of  model 2  becomes that of the model 1.  The difference between two models are the spin of the fields.
    
The partition function is defined by
\bel
\mathcal{Z}_{\text{CSM2}} (S^1 \times \Sigma_h) 
=\frac{1}{\text{Vol}(\mathcal{G})}\int \mathcal{D}^3 A_{\mu} \mathcal{D}^2 \lambda_{z}  \mathcal{D}^2 \phi_{i}  \mathcal{D}^2 \psi_{i} \mathcal{D}^2 Y \mathcal{D}^2 \chi
  e^{S_{\text{coh}}+ S_{\text{mat}2}} 
\ee 
The evaluation of the partition function is quite parallel to the first twist. Then the partition function is given by
\bel
&&\mathcal{Z}_{\text{CSM2}}(S^1 \times \Sigma_h) 
=(1- e^{-2 \pi m})^{-N(h-1)} 
\non && ~~~
\times \sum_{ \{ \varphi^a \}^{N}_{a=1} \in \mathcal{B}_2}  
  \left( \Bigl|  \text{det}_{c,d} \left[  \frac{\partial \beta^c_2}{ \partial \varphi^d }  \right] \Bigr| \prod_{a \neq b }
 \frac{1 } { (e^{2\pi i \varphi_a} - e^{2\pi i \varphi_b})(e^{2\pi i \varphi_a} -e^{-2\pi m} e^{2\pi i \varphi_b} )} \right)^{h-1} \non    
\label{partition1}
\ee
Here $\{ \varphi^a\}^{N}_{a=1}$ come form constant mode of gauge field $A^a_t (z,\bar{z})$.
and $\mathcal{B}_2$ is the set of  the solutions of the following equations:
\bel
2\pi i (k+2N) \varphi^a +\sum_{b=1 \atop b \neq a}^N \log 
\frac{e^{2\pi i \varphi^a} -e^{2\pi i \varphi^b} e^{-2\pi m} }{e^{2\pi i \varphi^a}  e^{-2\pi m}-e^{2\pi i \varphi^b} } = 2\pi i n^a, \quad n^a \in \Z.
\label{bethe2}
\ee

%%%%%%%%%%%%%%%%%%%%%%%%%%%%%%%%%%%%%%%%%%%%%
\subsection{ The correspondence of Correlation functions}
\label{correlation}
In this section we generalize the correspondence between partition function to  correlation functions of  $Q$-closed operator. 
The $Q$-closed gauge invariant operators in the  Chern-Simons-matter theory  are Wilson loops along the trivial fiber direction
\bel
W_{\nu} := \mathrm{Tr}_{\nu} \text{P} \exp \left(i \int_{S^1} A_t d t  \right).
\ee
Here   the  trace $\text{Tr}_{\nu}$ is taken  over a representation $\nu$ of gauge group.
A normalized correlation function of Wilson loops on genus-$h$ Riemann surface is defined by
\bel
\langle \prod_{l=1}^n W_{\nu_l}  \rangle &&:= 
  \frac{1}{\mathcal{Z}_{\text{CSM1}} (S^1 \times \Sigma_h) \text{Vol}(\mathcal{G})} \non
&& \times \int \mathcal{D}^3 A_{\mu} \mathcal{D}^2 \lambda_{i} \mathcal{D}^2 \phi   \mathcal{D}^2 \psi \mathcal{D}^2 Y_{i} \mathcal{D}^2 \chi_{i}
  e^{S_{\text{coh}}+ S_{\text{mat}}} 
 \prod_{l=1}^n W_{\nu_l} 
\ee

Again we take the gauge condition (\ref{gaugefix}). Then the Wilson loop operator is written as
\bel
W_{\nu} = \mathrm{Tr}_{\nu}  \exp \left( 2\pi i \sum_{a=1}^N A^a_t (z, \bar{z} ) H_a  \right)
\ee
Since the path integral over the component fields can be performed same way as partition function case.
The field configurations localized to  equation $\mathcal{B}_1$.
Then, the correlation function of Wilson loop in the model 1 is written as 
 \bel
&&\mathcal{Z}_{\text{CSM}} (\Sigma_h) \langle \prod_{l=1}^n W_{\nu_l}  \rangle 
= (1-e^{-2\pi m})^{N(h-1)}  \non
&& ~~~\times 
 \sum_{ \{ \varphi^a \}^{N}_{a=1} \in \mathcal{B}_1}  \prod_{l=1}^n  \mathrm{Tr}_{\nu_l} e^{2\pi i \sum_a \phi^a}   
  \left( \text{det}_{c,d} \left[  \frac{\partial \beta^c_1 (\varphi)}{ \partial \varphi^d } \right] \prod_{a \neq b }
 \frac{  e^{2\pi i \varphi_a} -e^{-2\pi m} e^{2\pi i \varphi_b} }
{e^{2\pi i \varphi_a} - e^{2\pi i \varphi_b}} \right)^{h-1} 
 \ee
As we mentioned before, the  over all  ambiguity does not affect the normalized correlation functions.
The corresponding objects in the $G/G$ gauged WZW-matter model is the correlation function of $G$-elements:
\bel  
\langle \prod_{l=1}^n \mathrm{Tr}_{\nu_l} g     \rangle, \quad (g : \Sigma_h \to G ).
\ee
Then The correlation function perfectly agree with 
\bel
\langle \prod_{l=1}^n W_{\nu_l}  \rangle
= \langle \prod_{l=1}^n \mathrm{Tr}_{\nu_l} g    \rangle 
\ee
We comment on the physical meaning of the above correlation in the $q$-boson model side. We showed that correlations function of
$G$-elements is same as the correlation functions of conserved charges in the $q$-boson model. See \cite{Okuda:2013fea} for the detailed identification .
Thus we have established that Gauge/Bethe correspondence of correlation functions
 between a Chern-Simons-matter theory (model 1) and q-boson model. 

%%%%%%%%%%%%%%%%%%%%%%%%%%%%%%%%%%%
\section{Reation to  moduli spaces }
\label{secindex}
\subsection{integration over moduli pace}
In this section we study the relation between   moduli spaces defined by particular equations and the partition functions of Chern-Simons-Matter theories.
%Before we consider the Chern-Simons-matter theories, 
 Recall that the partition function of $G/G$ gauged WZW model or equivalently pure Chern-Simons theory 
 have a nice interpretation as a geometric index over the moduli space of flat connection on the Riemann surface and 
it reduces to the volume of flat connection BF theory limit in \cite{Witten:1991we}.
 From this analogy, we conjectured that the partition function of  $G/G$ gauged WZW-matter model  is related to 
index over moduli space on $\Sigma_h$ of following equation in \cite{Okuda:2013fea}.
\bel
\mathcal{M}_1 = \left\{ (\overline{\phi}, \phi, A_i) \Big| \ep^{i j} F_{i j}+ \omega [\phi, \overline{\phi}]=0, D_{z} \phi=0, D_{\bar{z}} \overline{\phi}=0 \right\}/\mathcal{G}
\label{modulisp1}
\ee
Here $\omega=\sqrt{\text{det} (g_{\Sigma_h}})$ with the metric $g_{\Sigma_h}$ on Riemann surface  
Note that $(\overline{\phi}, \phi, A_i)$ in (\ref{modulisp1}) are two dimensioanl fields on the Riemann surface
 and do not depend on the time direction.
When  the BF theory-like limit is taken, the $G/G$ gauged WZW-matter model reduces to A-twisted gauged linear sigma model (GLSM) 
with an adjoint chiral multiplet. The partition function of the GLSM is interpreted as equivariant volume in the sense of \cite{Moore:1997dj}.   
 we can show path integral of the GLSM localize to the moduli space $\mathcal{M}_1$ by changing a $Q$-exact matter action.
 On the other hand, in level of  the gauged WZW-matter model, it was unclear 
how the path integral of gauged WZW-matter model localized to $\mathcal{M}_1$.
%But, it was unclear how  the partition function is interpreted as integration over the $\mathcal{M}_1$.. 
we will a give field theoretic explanation in terms of Chern-Simons-matter theories. 
%First we explain a physical reason why the partition function of Chern-Simons-Matter theory can be interpreted as a integration over moduli space.
In order to do this, we modify the previous $Q$-exact matter action as    
\bel
S_{\text{mat}1}^{\tau_0, \tau_1, \tau_2} && = \int_{S^1 \times \Sigma_h} Q \mathrm{Tr} 
\Bigl[ \tau_0  \overline{\phi}  \psi  + \tau_1 g^{z \bar{z}} Y_{{z}}  \chi_{\bar{z}} 
+ \tau_2 (D_z \phi \cdot \chi_{\bar{z}} +D_{\bar{z}} \overline{\phi} \cdot \chi_z ) \Bigr] 
\ee
with $\tau_i \in \C$.
Since the action is $Q$-exact, the partition function is independent of the generic  point in the parameter space $(\tau_0, \tau_1, \tau_2) $ at least the action is bound from below. 
In the slice $(\tau_0, \tau_1,\tau_2)=(1, 1,0)$, the action recover the original action (\ref{matteraction1}).
 On the other hand, the matter action is written in the slice 
 $(\tau_0, \tau_1,\tau_2)=(\frac{k}{4\pi} ,0,i)$ as
\bel
S_{\text{mat}1}^{\tau_0=-\frac{k}{4\pi}, \tau_1=0, \tau_2=i} &&=\int_{S^1 \times \Sigma_h} \mathrm{Tr} \Bigl[  \frac{k}{4\pi} 
\overline{\phi}  (D_t \phi + m \phi ) 
+ i g^{z \bar{z}} (D_{z} \phi Y_{\bar{z}} +D_{\bar{z}} \overline{\phi} Y_{z})+\cdots
%+i[\lambda_{z}, \phi] \chi_{\bar{z}})  
 %  -i g^{z \bar{z}} {\chi}_z  (D_t \chi_{\bar{z}}  - i m \chi_{\bar{z}})
  \Bigr]
\ee
Here ellipses indicate the $A_{t}, Y_z$ and $Y_{\bar{z}}$ independent terms. 
%When we perform the dimensional reduction along the $S^1$-direction, the partition function of the second twist reduces to...  
Then the path integral over the $A_{t}$, $Y_{\bar{z}}$ and $Y_{\bar{z}}$  impose the delta function constraints 
$\delta ( \omega [\phi, \overline{\phi}]+ \ep^{i j}F_{i j})$,
 $\delta (D_{z} \phi )$ and $\delta (D_{\bar{z}} \bar{\phi} )$, respectively. The partition function is written as
\bel
\mathcal{Z}_{\text{CSM1}} (S^1 \times \Sigma_h) 
&&=\int \mathcal{D}^2 A_{i} \mathcal{D}^2 \lambda_{i} \mathcal{D}^2 \phi   \mathcal{D}^2 \psi   \mathcal{D}^2 \chi_{i} 
\delta ( \omega [\phi, \overline{\phi}]+ \ep^{ i j} F_{i j}) \delta (D_{z} \phi ) \delta (D_{\bar{z}} \overline{\phi} ) \non
&&~~~~ \times  \exp \left( \int _{S^1 \times \Sigma_h} d^3 x \mathrm{Tr} 
\Bigl[ \frac{ik}{4 \pi} \ep^{ij} A_i \partial_t A_{j} -\frac{k}{4\pi} \sqrt{g_{(3)}} , \overline{\phi} \partial_{t} \phi  \cdots \Bigr] \right) 
  \label{eff2}
\ee 
Thus we find that the path integrals  localize to the moduli space (\ref{modulisp1}). 
After we perform the path integrals for the field configurations $\prod_{z, \bar{z} \in \Sigma_h} d \Psi(t, z,\bar{z})$. 
$\mathcal{Z}_{\text{CSM1}} (S^1 \times \Sigma_h)$ is interpreted the partition function of effective quantum mechanics on $S^1$ whose target space is $ \mathcal{M}_1$.
%\bel
%S_{\text{mat}2, \tau_1=0, \tau_2=1} &&=\int_{S^1 \times \Sigma_h} \mathrm{Tr} \Bigl[   \frac{k}{4 \pi} {\phi}_{z}  (D_t \phi_{\bar{z}} + m \phi_{\bar{z}} ) 
%+ i (g^{z \bar{z}}(D_{z} \phi_{\bar{z}} Y + g^{\bar{z} z}D_{\bar{z}} {\phi}_{z} )+\cdots
%+i[\lambda_{z}, \phi] \chi_{\bar{z}})  
 %  -i g^{z \bar{z}} {\chi}_z  (D_t \chi_{\bar{z}}  - i m \chi_{\bar{z}})
  %\Bigr]
%\ee 
Similarly, in the model 2,  we can show that  the path integral localizes to the moduli space of Hitchin equation \cite{Hitchin:1986vp}:
\bel
\mathcal{M}_{\text{Hitchin}} = 
\left\{ ({\phi}_{z}, \phi_{\bar{z}}, A_i) \Big| F_{z \bar{z}}+  [{\phi}_{z}, \phi_{\bar{z}}]=0, D_{z} \phi_{\bar{z}}=0, D_{\bar{z}} {\phi}_{z}=0 \right\}/\mathcal{G}
\label{modulisp2}
\ee

 In the pure Chern-Simons theory case, 
 quantization of Chern-Simons theory corresponds to quantization of the moduli space of flat connection.
We expect that the quantization of twisted Chern-Simons-matter theory is related to quantization of $\mathcal{M}_1 $ or $\mathcal{M}_{\text{Hitchin}}$.
 Here  $A_i \partial_t A_{j}$ and $\overline{\phi} \partial_{t} \phi$  behaves as
 the kinetic term $p_i \dot{q}_j$ in the Hamiltonian formalism of effective quantum mechanics.
It is interesting to study  quantization of moduli spaces  and its relation to $q$-boson model. This problem is left to our future work. 
Again, the partition function of model 2 define the effective quantum mechanics whose target space is the moduli space of Hitchin equation on Riemann
surface:

\subsection{index over moduli space: $SU(2)$ case}

We conjectured in \cite{Okuda:2013fea} that the partition function of $G/G$ gauged WZW-matter model is related to a geometric index \cite{Teleman:2003}. In this subsection, we perform the identification of 
the partition function and index for $G=SU(2)$ case.
  
The moduli space $\mathfrak{M}$ relevant to $G/G$ gauged WZW-matter model is roughly speaking the isomorphism classes of holomorphic $G_{\C}$-bundle on genus $h$ Riemann surface
 and the bundle on $\mathfrak{M}$ is
\bel
\lambda_t (T \mathfrak{M}) \otimes  \mathcal{O}(k) \otimes E^{*}_x U
\ee  
Here $\lambda_t (T \mathfrak{M})$ is defined by $K^0$-classes $[\wedge^n T \mathfrak{M}]$ of anti-symmetric products of the tangent bundle $\wedge^n T \mathfrak{M}$ as:
\bel
\lambda_t (T \mathfrak{M}) :=1+\sum_{n=1}^{\infty} (-t)^n  [\wedge^n T \mathfrak{M}] \in K^{0} (T\mathfrak{M};\mathbb{Q})[t],
\ee
$\mathcal{O}(k)$ is the line bundle with chern class $k$. % and $E^{*}_x U$ is

For simplicity, we consider the $G=SU(2)$ case, The general cases will be studied in our future work.
The index $\text{Ind}(\mathfrak{M}, \lambda_t (T \mathfrak{M}) \otimes  \mathcal{O}(k) \otimes E^{*}_x U) $ is given by
\bel
&&\text{Ind}(\mathfrak{M}, \lambda_t (T \mathfrak{M}) \otimes  \mathcal{O}(k) \otimes E^{*}_x U) \non
&&~~~=\sum_{u} (1-t)^{h-1} \left[ \frac{(1-tu^2)(1-t u^{-2})}{(u-u^{-1})^2} \right]^{h-1}  \left[2k+4 +4 t \left(\frac{u^2}{1-tu^2} + \frac{u^{-2}}{1-tu^{-2}}\right) \right]^{h-1} \non
\ee
Here $\sum_u $ runs over the set of solution of following equation
\bel
u^{2k+4} \left[ \frac{1-t u^{-2}}{1-t u^2}\right]^2 =1.
\label{constrsu}
\ee 
Next we calculate the partition function of $SU(2)/SU(2)$ gauged WZW-matter model or equivalently the partition function of model 1, 
the ratio of functional determinants of gauge field and ghost field becomes
\bel
\prod_{\alpha \neq 0} \frac{\mathrm{Det}_{c} (\partial_t +i \alpha(A_t) )}{\mathrm{Det}_{A_i} (\partial_t +i \alpha(A_t) )}
= \exp \Bigl\{  \frac{1}{8\pi} \int_{\Sigma_h} R \log  \left( 1-e^{4 \pi i  A_t} \right) \left( 1-e^{-4 \pi i A_t} \right)   
+\frac{1}{2\pi} \int_{\Sigma_h}   F \log \frac{ 1-e^{4 \pi i A_t}}{ 1-e^{-4 \pi i  A_t}}   \Bigr\}. \non
\ee
The ratio of functional determinants of matter  fields becomes
\bel
&&\frac{\mathrm{Det}_{\chi} (\partial_t +i \alpha(A_t) - m)}{\mathrm{Det}_{\phi} (\partial_t +i \alpha(A_t) - m)} =(1-e^{-2\pi m}) \non
&&~~\times\exp \Bigl\{  -\frac{1}{8\pi} \int_{\Sigma_h} R \log  \left( 1-e^{4 \pi i  A_t-2\pi m} \right) \left( 1-e^{-4 \pi i A_t-2 \pi m} \right)   
-\frac{1}{2\pi} \int_{\Sigma_h}   F \log \frac{ 1-e^{4 \pi i A_t-2\pi m}}{ 1-e^{-4 \pi i  A_t-2\pi m}}   \Bigr\}. \non 
\ee
The $\beta^a(\varphi)_1$ becomes
\bel
\beta^a(\varphi)_1&&= \frac{1}{2\pi } \Bigl[2 \pi i k \varphi 
+\log \left( \frac{1-e^{4 \pi i \varphi }}{1-e^{-4 \pi i \varphi }}\right) 
+ \log \left( \frac{1-e^{4 \pi i \varphi -2 \pi m }}{1-e^{-4 \pi i \varphi -2 \pi m }}\right) \Bigr] \non
&&=\frac{1}{2\pi }  \Bigl[2 \pi i ( k+2 ) \varphi +{\pi i} 
+ 2 \log \left( \frac{1-e^{4 \pi i \varphi -2 \pi m }}{1-e^{-4 \pi i \varphi -2 \pi m }}\right) \Bigr]
\ee
Then partial integration and using poisson resummation formula impose the following field configuration
 \bel
2 \pi i ( k+2 ) \varphi 
+ 2 \pi i \log \left( \frac{1-e^{4 \pi i \varphi -2 \pi m }}{1-e^{-4 \pi i \varphi -2 \pi m }}\right) \in 2 \pi i (n+\frac{1}{2}), \quad n \in \Z 
\ee
and the partion function of $SU(2)/SU(2)$ gauged WZW-matter model is given by
 \bel
\frac{\partial \beta_1(\varphi)}{\partial \varphi}&&=i
\left[ k+2 +2 e^{-2\pi m} \left( \frac{e^{-4 \pi i \varphi -2 \pi m}}{1-e^{-4 \pi i \varphi -2 \pi m}} +\frac{e^{4 \pi i \varphi -2 \pi m}}{1-e^{4 \pi i \varphi -2 \pi m}} \right)\right] 
%&&=i \left[k+2 +2 t \left( \frac{t u^{-2} }{1-t u^{-2}} + \frac{t u^{2} }{1-t u^{2}}\right) \right]
\ee
Then the partition function is given by
\bel
&&\mathcal{Z}^{\text{SU(2)}}_{\text{CSM}}(S^1 \times \Sigma_h)=(1-e^{-2\pi m})^{h-1} \sum_{x \in \mathcal{B}_1}
\left(\frac{(1-e^{-2\pi m} e^{-4\pi i \varphi })(1-e^{-2\pi m} e^{4\pi i \varphi })}{(e^{2\pi i \varphi }-e^{-2\pi i \varphi })^2  } \right)^{h-1} \non
&&~~~~~~~~~~~~~~~~ \times \left[ 2k+4 +4 e^{-2\pi m} \left( \frac{e^{-4 \pi i \varphi -2 \pi m}}{1-e^{-4 \pi i \varphi -2 \pi m}} +\frac{e^{4 \pi i \varphi -2 \pi m}}{1-e^{4 \pi i \varphi -2 \pi m}} \right)\right]^{h-1} 
\ee
When we define $u:=e^{2\pi i\varphi}$ and $t:=e^{-2\pi m}$,  $\mathcal{Z}^{\text{SU(2)}}_{\text{CSM}}(S^1 \times \Sigma_h)$ agrees with (\ref{constrsu}) up to a overall factor.
Thus the partition function has the interpretation of index.

%%%%%%%%%%%%%%%%%%%%%%%%%%%%%%%%%%%%%%%%%%%%%%%%%%%%%%%%%%%%%%%%%%%%%%%%%%
\section{Summary}
\label{summary}

In this paper we evaluated the partition functions of twisted Chern-Simons-matter theory on $S^1 \times \Sigma_h$ and showed that the partition function agrees 
with the partition functions of  $G/G$ gauged WZW-matter model on $\Sigma_h$.
We revealed  Chern-Simons-matter theory on $S^1 \times \Sigma_h$  also possess hidden quantum integrable structure . 
We also evaluated correlations of Wilson loops in twisted Chern-Simons-matter theory and showed that these correlation functions perfectly agree with the correlation functions 
of $G$-elements in a generalizations of  $G/G$ gauged WZW-model.    

The partition function of the model 1 is also expressed by norms of eigen states of $q$-boson model and have  TQFT structure.
More over we explicitly showed that the partition function is can be interpreted as an index over moduli space of $G$-bundles for $SU(2)$ case. 

 On the other hand, the physical interpretation of the partition function of model 2 
 is not clear in the quantum integrable side. 
 It is interesting to find corresponding object of the partition function of model 2 in the quantum integrable model side.
us

%%%%%%%%%%%%%%%%%%%%%%%%%%%%%%%%%%%%%%%%%
%%%%%%%%%%%%%%%%%%%%%%%%%%%%%%%%%%%%%%%%
\subsubsection*{Note added}
When this work was being completed, there appeared a paper \cite{Gukov:2015sna} which
has  substantial overlap with ours.
%%%%%%%%%%%%%%%%%%%%%%%%%%%%%%%%%%%%%%%%%%%
%%%%%%%%%%%%%%%%%%%%%%%%%%%%%%%%%%%%%%%%%%%%%%  

\subsection*{Acknowledgment}
YY is grateful to Kazushi Ueda for explanation  of \cite{Teleman:2003} and many interesting discussions. 
He is also grateful to Daisuke Yokoyama for kind  hospitality during 
 The 2nd Workshop on Developments in M-Thoery at High 1 Resort, Gangwondo, Korea.
%%%%%%%%%%%%%%%%%%%%%%%
%%%%%%%%%%%%%%%%%%%%%%%
%%%%%%%%%%%%%%%%%%%%%%%
%%%%%%%%%%%%%%%%%%%%%%%%%%%%
%\newpage

\end{document}